\documentclass[conference]{IEEEtran}
\IEEEoverridecommandlockouts
\usepackage{cite}
\usepackage{amsmath,amssymb,amsfonts}
\usepackage{algorithmicx}
\usepackage{algorithm}
\usepackage{graphicx}
\usepackage{textcomp}
\usepackage{xcolor}
\usepackage{cite}  
\usepackage{amsmath,amssymb,amsfonts}
\usepackage{textcomp}
\usepackage{xcolor}    
\usepackage{amsthm}  
\usepackage{graphicx}  
\usepackage{multirow}  
\usepackage{algpseudocode}
\usepackage{subfigure}
\usepackage{ulem}

\def\BibTeX{{\rm B\kern-.05em{\sc i\kern-.025em b}\kern-.08em
    T\kern-.1667em\lower.7ex\hbox{E}\kern-.125emX}}
\begin{document}

\title{Securing Semantic Communications with Physical-layer Semantic Encryption and Obfuscation\\
{\footnotesize \textsuperscript{}}
\thanks{* Guoshun Nan is the corresponding author.} 
}

\author{
	\IEEEauthorblockN{
		Qi Qin,
		Yankai Rong,
  		Guoshun Nan*,
		Shaokang Wu,
		Xuefei Zhang,
		Qimei Cui,
		Xiaofeng Tao}
	\IEEEauthorblockA{Beijing University of Posts and Telecommunications}
	\IEEEauthorblockA{q\_qin@bupt.edu.cn, yankairong@bupt.edu.cn,  nanguo2021@bupt.edu.cn, w\_s\_k@bupt.edu.cn, \\ zhangxuefei@bupt.edu.cn,  cuiqimei@bupt.edu.cn, taoxf@bupt.edu.cn}
}

\maketitle

\begin{abstract}
Deep learning based semantic communication (DLSC) systems have shown great potential of making wireless networks significantly more efficient by only transmitting the semantics of the data. However, the open nature of wireless channel and fragileness of neural models cause DLSC systems extremely vulnerable to various attacks. Traditional wireless physical layer key (PLK), which relies on reciprocal channel and randomness characteristics between two legitimate users, holds the promise of securing DLSC. The main challenge lies in generating secret keys in the static environment with ultra-low/zero rate. Different from prior efforts that use relays or reconfigurable intelligent surfaces (RIS) to manipulate wireless channels, this paper proposes a novel physical layer semantic encryption scheme by exploring the randomness of bilingual evaluation understudy (BLEU) scores in the field of machine translation, and additionally presents a novel semantic obfuscation mechanism to provide further physical layer protections. Specifically, 1) we calculate the BLEU scores and corresponding weights of the DLSC system. Then, we generate semantic keys (SKey) by feeding the weighted sum of the scores into a hash function. 2) Equipped with the SKey, our proposed subcarrier obfuscation is able to further secure semantic communications with a dynamic dummy data insertion mechanism. Experiments show the effectiveness of our method, especially in the static wireless environment.    

\end{abstract}


\section{Introduction}
Deep learning-based semantic communication (DLSC) has shown great potentials in boosting communication efficiency by merging one or multiple modules in the traditional block-wise system. It transmits semantics conveyed by the source rather than the accurate reception of each single symbol or bit. Such a novel communication paradigm significantly facilitates newly emerging mobile applications which require high throughput and low latency, such as VR/AR, and Human-to-Human communications. Specifically, the transmitter of a DLSC system\cite{b2} relies on neural networks to extract semantic information from the input and then sends encoded symbols to wireless channels. Next, the receiver reconstructs the input based on the semantics from the sender with neural models or interprets the semantics to make intelligent decisions.


Although promising, DLSC systems may suffer from more challenging security issues compared with traditional wireless communication due to two reasons as follows. The open nature of wireless channels leads to a high risk of malicious attacks against semantics confidentiality in DLSC. Meanwhile, DLSC systems are more vulnerable to eavesdropping, tampering and spoofing because of the fragileness of deep neural networks. Previous works proposed for block-wise communication systems secure messages using either symmetric-key or asymmetric-key encryption schemes\cite{b12}. To reduce the computation and communication overhead, physical layer key (PLK) schemes \cite{b13,b14} explore the reciprocity principle and temporal variation of the wireless channels between two legitimate users to yield efficient encryption and decryption. From the perspective of information theoretical security, PLK also enables wireless communications to achieve Shannon’s perfect encryption.
However, as temporal decorrelations of wireless channels may not be always guaranteed, the secret key generation rates tend to be ultra-low/zero in same static fading environments such as indoor IoT networks.

Previous PLK works \cite{b17} mainly employ additional relays or reconfigurable intelligent surfaces (RIS) to inject randomness of wireless channel by manipulating signals in static environments. Different from these efforts, this paper proposes a novel physical layer semantic encryption/decryption scheme by exploring the randomness of BLEU (Bilingual evaluation understudy) scores\cite{b7} in the field of natural language processing, where BLEU is a standard benchmark for evaluating the machine translations against the human translations. We further present a semantic obfuscation mechanism to provide further protection. We use orthogonal frequency-division multiplexing (OFDM)\cite{b18} to transmit the underlying semantics as it dominates existing cellular communications. Prior studies\cite{b18} also discussed the promise of PLK in OFDM systems, although uplink and downlink fading are different. 


Specifically, our method consists of two key steps, i.e., semantic encryption and semantic obfuscation. 1) In the first step, we generate semantic secret keys (SKey) to encrypt semantic data by computing BLEU scores and the corresponding weights in 1-gram, 2-gram, 3-gram and 4-gram settings, where N-gram here indicates the evaluation of just matching grams of a specific order, such as single words (1-gram) or word pairs (2-gram). Our encryption method naturally involves the randomness during key generation because of the deviations of semantics, so that the static fading environment issue can be properly alleviated. 2) To make the encrypted OFDM symbols more random and secure, we group $s$ OFDM symbols as a data unit and also reserve $k$ subcarriers for dummy data, which are denoted as [$s$, $k$]. We then obfuscate them by inserting dummy data in between the encrypted semantic data. Different from the prior work\cite{b1} that fixed $k$ and $s$, the two values in our paper are computed by SKeys and vary from each group of OFDM symbols to induce more randomness for the obfuscation. Specifically, the contributions of this paper can be summarized as follows: 

\begin{itemize}
\item We present a novel semantic key generation method by exploring the BLEU scores in the field of machine translation to secure the semantic communications, facilitating the key generation in static fading environments. 
\item We introduce a novel physical layer semantic encryption and semantic obfuscation method, where the former takes the semantic key as input to encrypt the semantic information, and the latter carries out subcarrier-level obfuscation by a dynamic dummy data insertion mechanism. 
\item We conduct experiments to show the effectiveness of our approach and further provide quantitative analyses based on the search space to verify the security of the encryption and obfuscation method.
\end{itemize}

\section{Related Work}
\subsection{Semantic Communications}

Deep learning based semantic communications (DLSC) \cite{b2,b9,b10,b16,b19,b20} have been proven effective for data transmission over wireless networks. However, the research of DLSC security is still on an early stage. Previous works along this line can be categorized into privacy protection \cite{b11} and system's robustness \cite{b15}. The most related work is SIoT which considers physical layer security of DLSC in the IoT \cite{b4} environment. Our work differs from the above efforts in two aspects: 1) our method aims to generate semantic keys by exploring the BLEU scores in the field of machine translation, and such a method can be applied to static fading environments. While the previous work only analyzes  traditional physical key generation methods in DLSC. 2) we additionally present a novel OFDM subcarrier obfuscation method to further secure the communication.

\subsection{OFDM Physical Layer Encryption}
The research on OFDM system can be traced back many years ago in \cite{b8}. Early study \cite{b3} proposed a chaos based OFDM constellation scrambling method for physical layer security of wireless communications. Towards this direction, more recent works \cite{b5,b6} aim to secure OFDM systems using scrambling or interleaving techniques. Different from the above works, we rely on semantic keys to obfuscate the OFDM symbols.

\section{Principle}

\subsection{Overview}
\begin{figure*}[htbp]
\centerline{\includegraphics[width=1\linewidth]{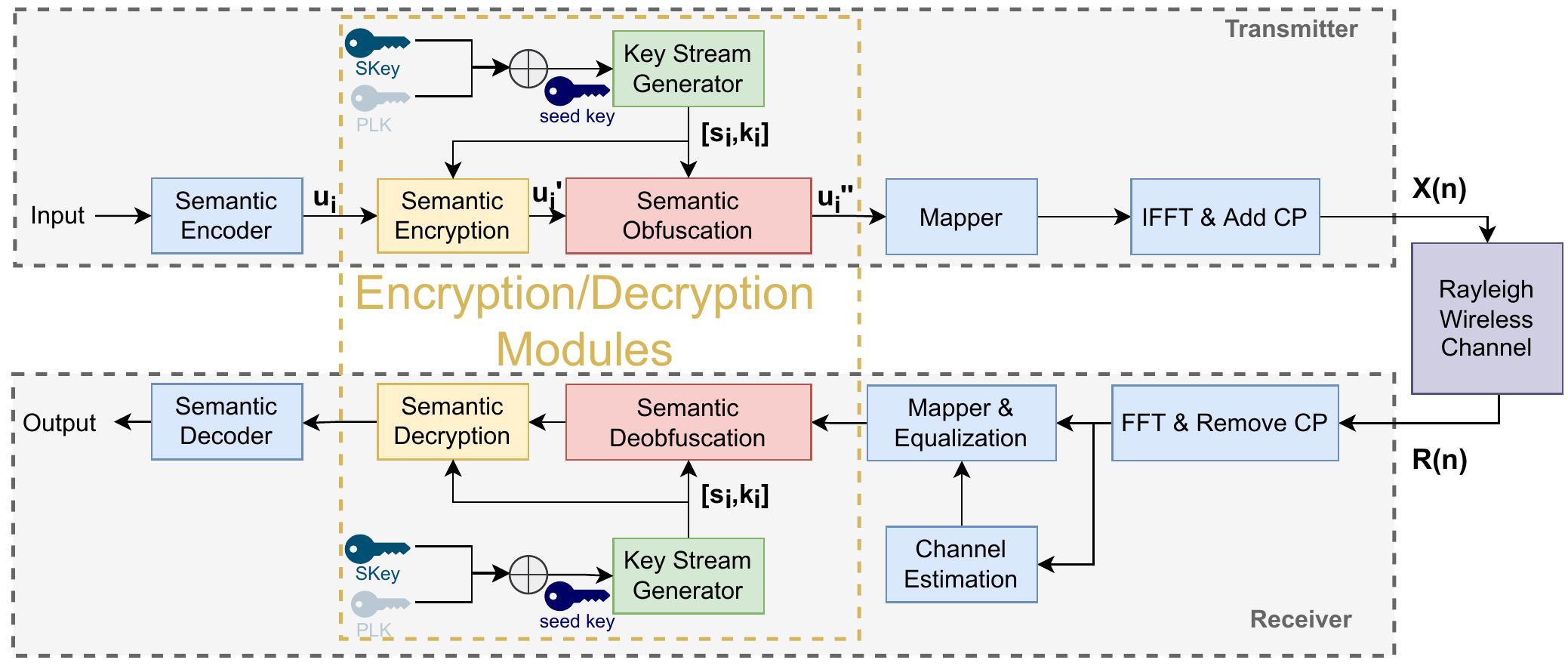}}
\vspace{-4mm}
\caption{Our proposed encryption and obfuscation methods. The former relies on semantic keys to encrypt the semantic representation, where the keys are generated by exploring BLEU scores. The latter utilizes our proposed subcarrier obfuscation mechanism to further secure semantic communications.}
\label{fig:arc}
\end{figure*}

In this section, we present our proposed semantic encryption and obfuscation methods for DLSC. Fig. \ref{fig:arc} shows the architecture of a DLSC system equipped with our two modules. Our semantic encryption module takes the semantic representation $\mathbf{u}_i$ of the $i$-th data as an input, where $\mathbf{u}_i \in \mathbb{R}^{d}$ is generated by the semantic encoder. Here $d$ is the dimension. The module encrypts $\mathbf{u}_i$ to output $\mathbf{u'}_i \in \mathbb{R}^{d}$ by exploring the semantic deviation of the DLSC system. Then, we feed the encrypted $\mathbf{u'}_i$ into our subcarrier obfuscation module to further secure the communication, where the output can be expressed as $\mathbf{u''_i} \in \mathbb{R}^{d}$. Finally, we feed $\mathbf{u''_i}$ into the OFDM systems to obtain a complex data point $\mathbf{X}(n)$\cite{b1}, which will be sent to wireless channel. The received signal $\mathbf{R}(n)$ in the frequency domain can be expressed as
\begin{equation}
R(n) = \mathbf{X}(n)\mathbf{H}(n) + \mathbf{N}(n),
\end{equation}
where $\mathbf{H}(n)$ and $\mathbf{N}(n)$ indicate the channel effect and noise effect, respectively \cite{b1}. The receiver takes ${R}(n)$ as input for demodulation, semantic deobfuscation, decryption and interpretation. Next we delve into the details of our methods.

\begin{figure}[htbp]
\centerline{\includegraphics[width=1\linewidth]{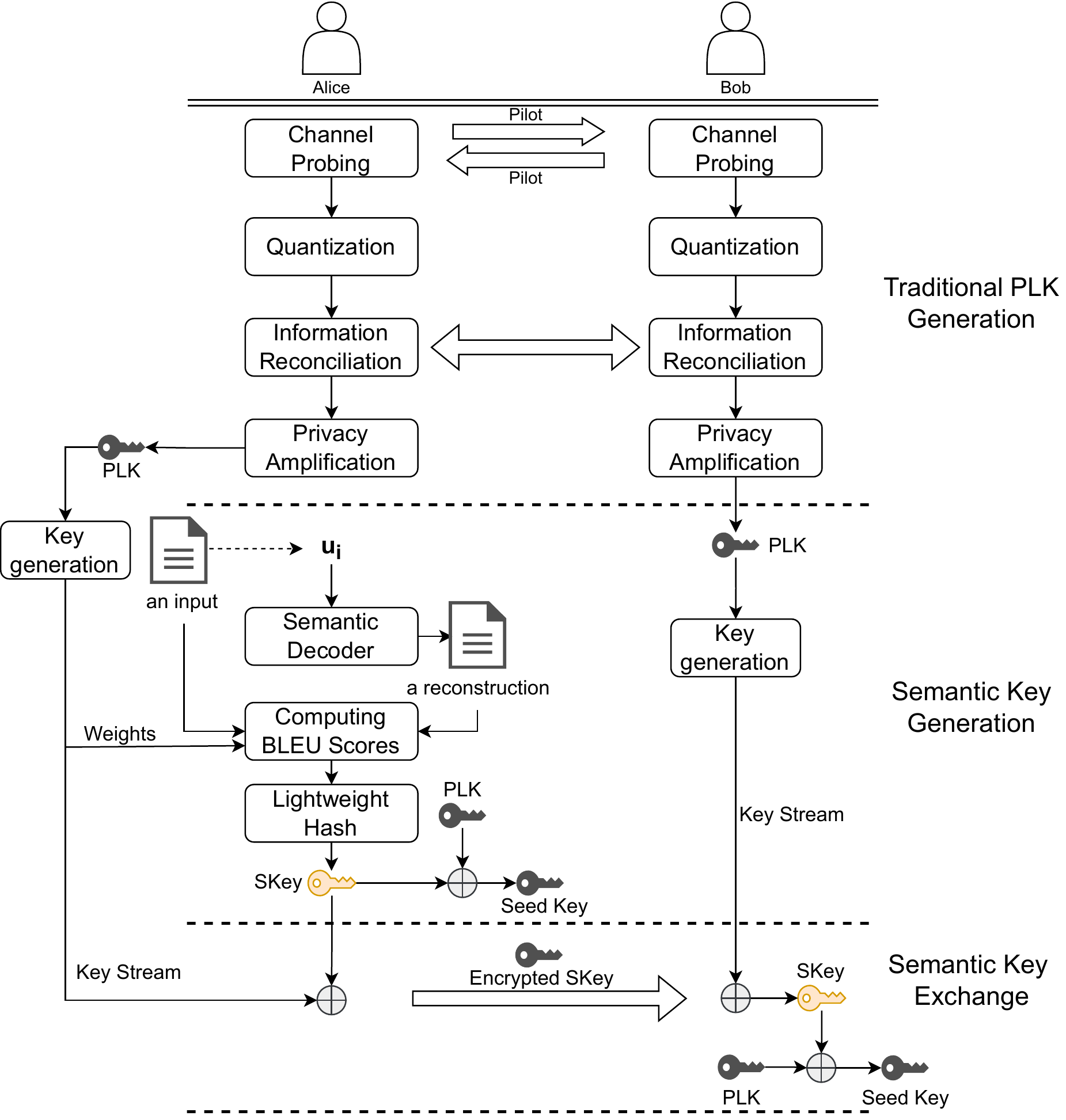}}
\caption{The procedure of semantic key generation and exchange. We feed the $i$-th data to the semantic to output the semantic representation $\mathbf{u}$, which can be used to compute the BLEU scores. Then we rely on the above BLEU scores and traditional PLK to generate the semantic key SKey.}
\label{fig:skey}
\end{figure}

\subsection{Semantic Encryption}

\noindent
\textbf{Semantic Key Generation:} Fig. \ref{fig:skey} illustrates how we generate our semantic keys at the transmitter for the $i$-th data. We first produce traditional physical layer key (PLK) with four steps including channel probing, quantization, reconciliation and privacy amplification, and then generate key stream $k_{stream}$. We also obtain the various BLEU scores $\mathbf{s} = [s_{i,1}, s_{i,2}, s_{i,3}, s_{i,4}]$ under the $1-$gram, $2-$gram, $3-$gram, and $4-$gram settings, by feeding $\mathbf{u_i}$ to the semantic decoder deployed at the transmitter. We then compute the weights of each BLUE score with the key stream $k_{stream}$. Finally, we generate the semantic key SKey by feeding the weighted sum of BLUE scores $\mathbf{s}$. Such a procedure can be expressed as

\begin{equation}
SKey = HASH(\mathbf{W} \times BLEU(data_{real},data_{predicted})),
\end{equation}
where $\mathbf{W}$ indicates BLEU score weights, BLEU represents the calculation of BLEU scores, and $HASH$ refers to lightweight hash algorithm. Algorithm \ref{alg:key} describes the procedure of SKey generation. Once SKey is obtained, it will be encrypted by PLK and sent to the other side of the communication. The receiver only needs to use PLK to decrypt the encrypted SKey to obtain SKey, and thus share SKey to two end points.

\begin{algorithm}[h]
\caption{SKey generation}
\hspace*{0.02in} {\bf INPUT:} 
data$_{Real}$, $seed$(PLK or (SKey  $\oplus$ PLK  )) \\
\hspace*{0.02in} {\bf OUTPUT:} 
SKey

\begin{algorithmic}[1]
\State  data$_{Encoded}$ = Semantic\_Encoder(data) 
\State  dat$a_{Predicted}$ = Semantic\_Decoder(data$_{Encoded}$)
\State  1\_gram,\;2\_gram,\;3\_gram,\;4\_gram = BLEU(data$_{Real}$,\;data$_{Predicted}$)
\State  weight$_{1}$,\;weight$_{2}$,\;weight$_{3}$,\;weight$_{4}$ = Weight\_generator(seed)
\State  Generated\_BLEU = (1\_gram $\times$  weight$_{1}$ + 2\_gram $\times$  weight$_{2}$ + 3\_gram $\times$  weight$_{3}$ + 4\_gram $\times$  weight$_{4}$) $\pmod 1$ 
\State  SKey = Lightweight\_Hash(Generated\_BLEU)
\end{algorithmic}
\label{alg:key}
\end{algorithm}


\noindent
\textbf{Semantic Encryption:} So far we have generated the semantic key SKey, which can be used to encrypt $\mathbf{u}_i$ to produce $\mathbf{u'}_i$ in the semantic encryption module of the transmitter. We then obtain the seed key by the operation SKey $\oplus $ PLK to generate the encryption and decryption key stream. SKey will be further encrypted using the key stream generated from the traditional PLK, and then sent to the legitimate receiver. Such a key can be used for the subsequent decryption.

\subsection{Semantic Obfuscation}
\begin{figure}[htbp]
\centerline{\includegraphics[width=1\linewidth]{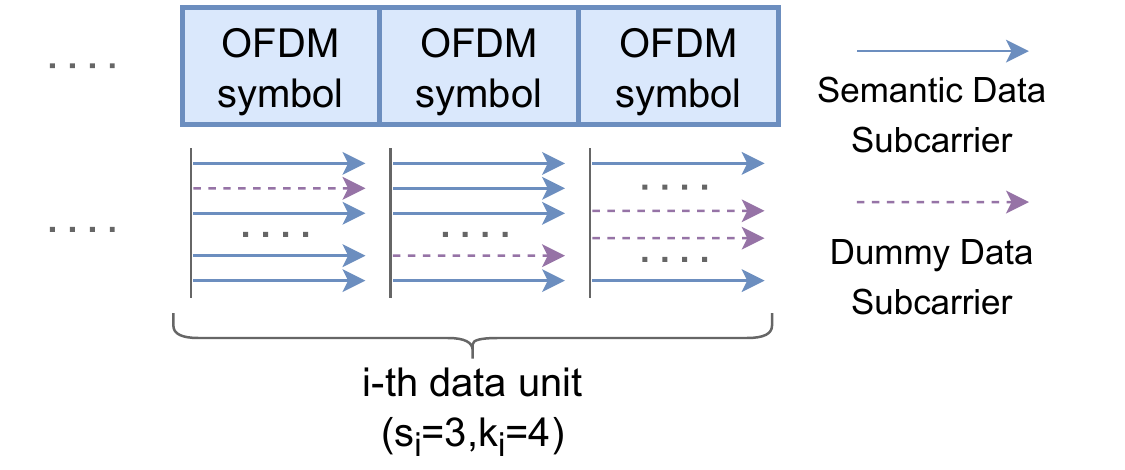}}
\vspace{-4mm}
\caption{Illustration of our semantic obfuscation. Here the number of OFDM symbols and dummy subcarrier, which are denoted as $s_{i}$ and $k_{i}$, are set as 3 and 4, respectively.} 
\label{fig:obfuscation}
\end{figure}

To further secure semantic communications, we introduce a novel subcarrier-level obfuscation technique to obfuscate the source by inserting dummy data between the encrypted data. The process is detailed as follows. In light of the previous effort \cite{b1}, we rely on the SKey and the traditional PLK to generate dummy data by our semantic encoder, and then we are able to obfuscate $\mathbf{u'}_i$ by inserting the above dummy data between encrypted semantic data. Fig. \ref{fig:obfuscation} illustrates the procedure of the above semantic obfuscation. 

As the source can be modulated based on multiple OFDM symbols, we group $s$ OFDM symbols into a data unit and reserve $k$ subcarriers for dummy data insertion. We denote such an obfuscation as a pair [$s$,$k$]\cite{b1}. 
Different from the prior work\cite{b1} that fixed $k$ and $s$, the two values in our approach are computed by SKey and vary from each group of OFDM symbols to induce more randomness for the obfuscation. We also denote the upper bound of $k$ and $s$ as $k_{max}$ and $s_{max}$ to trade off the transmission efficiency and the security performance. 

Algorithm \ref{alg:enc} describes the process of semantic encryption and subcarrier obfuscation. The character $b$ is the number of bits per subcarrier, which is determined by the mapping scheme, and $N_{d}$ refers to the number of the data subcarriers. At the receiver side, the deobfuscation module will remove dummy data in OFDM subcarriers and then feed the output to the semantic decryption. As we randomize the values of $s$ and $k$ for the above semantic obfuscation/deobfuscation, it will be much more difficult for eavesdroppers to decode the signal.

\begin{algorithm}[h]
\caption{Data encryption and subcarrier level semantic data obfuscation}
\hspace*{0.02in} {\bf Input:} 
data$_{bits}$, seed, [s$_{max}$, k$_{max}$],  b      \\
\hspace*{0.02in} {\bf Output:} 
data$_{enc}$

\begin{algorithmic}[1]
\State l$_{d}$ = length(data$_{bits}$)
\State key$_{xor}$ = key\_generator(seed,l$_{d}$)
\State data$_{xor}$ = data$_{bits}\oplus$ key

\While {(true)}
\State		s = int(key\_generator(seed))(mod$\enspace$s$_{max}$)  
\State		k = int(key\_generator(seed))(mod$\enspace$k$_{max}$) 
\State		if ((s $\times$ N$_{d}$ $ -$ k) $\times$ b) $>$ l$_{d}$ :
\State		$\quad\quad$break
\State		l$_{d}$ = l$_{d}$ - (s $\times$ N$_{d} $ $-$ k) $\times$ b
\State		\textit{DummySubcLoc} = Location\_generator(seed)
\State		\textit{DummyData} = 
\State		Semantic\_Encoder(key\_generator(seed2), kb)
\State		\textit{DataSubcLoc} = $\left [1:N_{d} \times s \right]$ - \textit{DummySubcLoc}
\State 		data$_{enc}$(\textit{DataSubcLoc}) = data$_{xor}$
\State 		data$_{enc}$(\textit{DummySubcLoc}) = \textit{DummyData}
\EndWhile

\end{algorithmic}
\label{alg:enc}
\end{algorithm}

\section{Simulation And Analysis}
We implement the DLSC system with 16QAM OFDM to verify the effectiveness of our methods. The number of OFDM subcarriers is set to 64 and we refer to DeepSC \cite{b2} for the implementation of our semantic encoder and decoder. 

\subsection{Comparisons of Bit Error Rate}
\begin{figure*}[htbp]
\begin{center}
\subfigure[The bit error rate (BER) as SNR change.]{
\label{fig:BER}
\includegraphics[width=2.25in,height=1.8in]{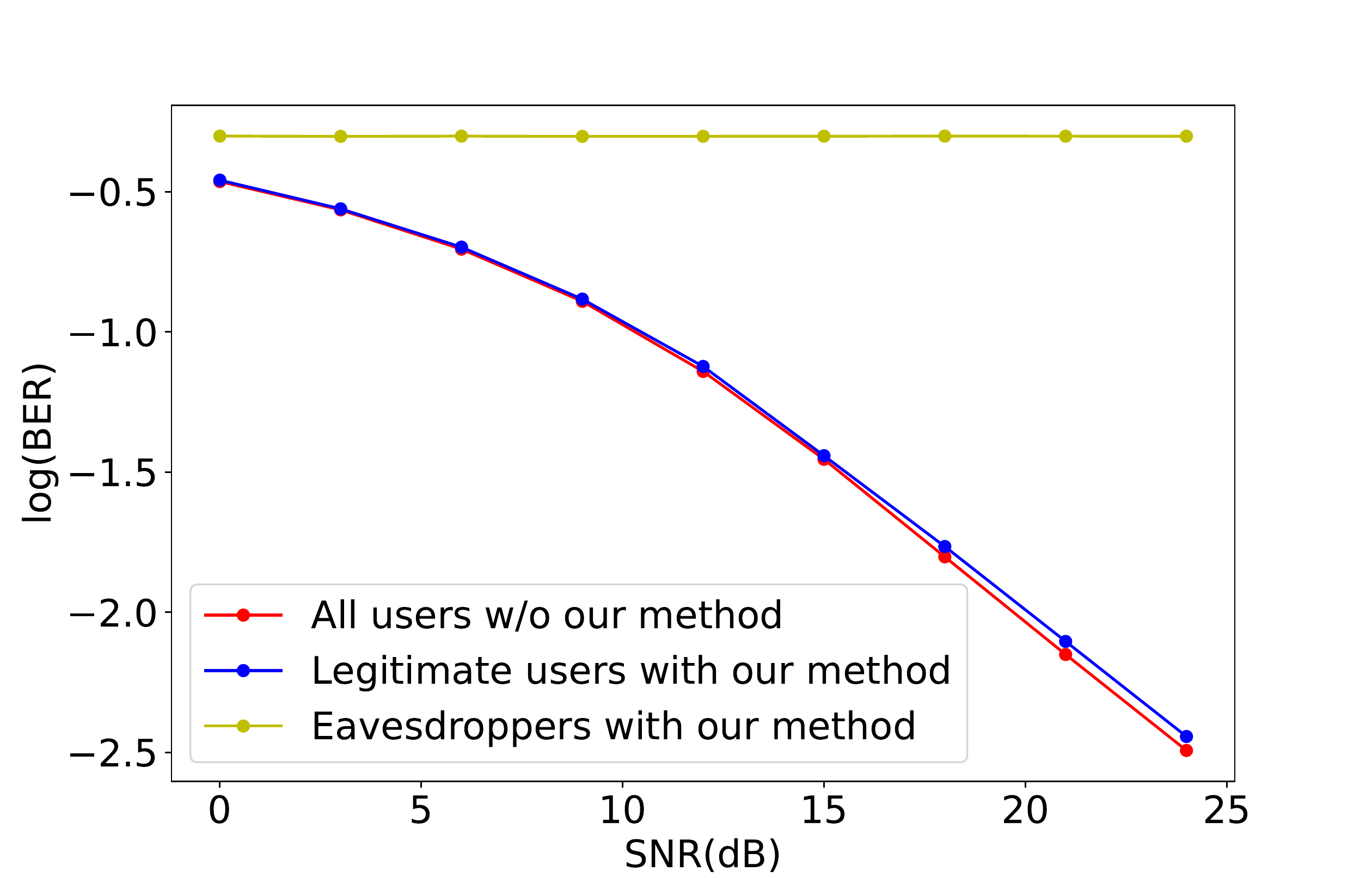}

}
\subfigure[BLEU comparisons under the 1-gram setting.]{
\label{fig:1g}
\includegraphics[width=2.25in,height=1.8in]{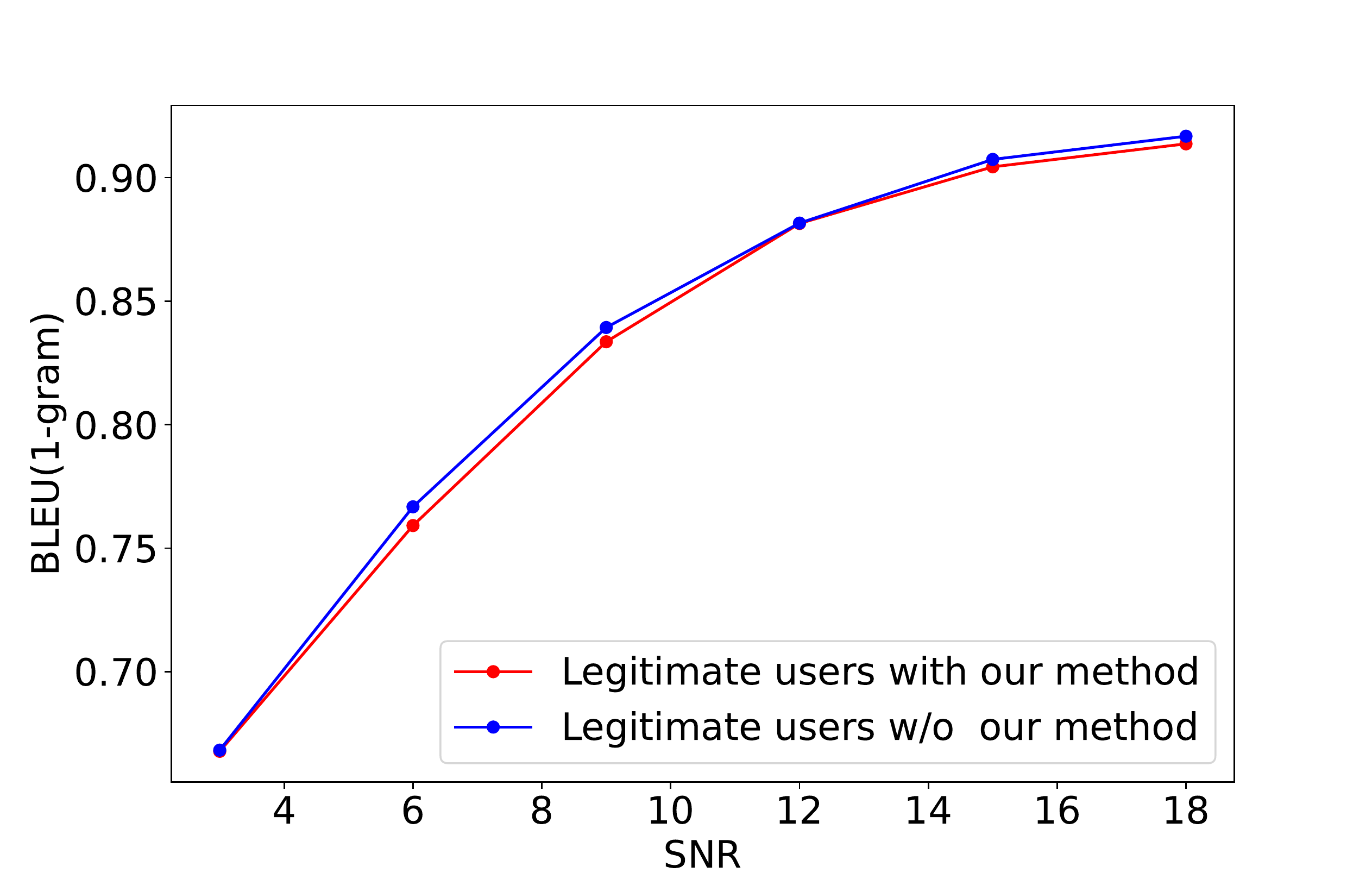}
}
\subfigure[BLEU comparisons under the 2-gram setting.]{
\label{fig:2g}
\includegraphics[width=2.25in,height=1.8in]{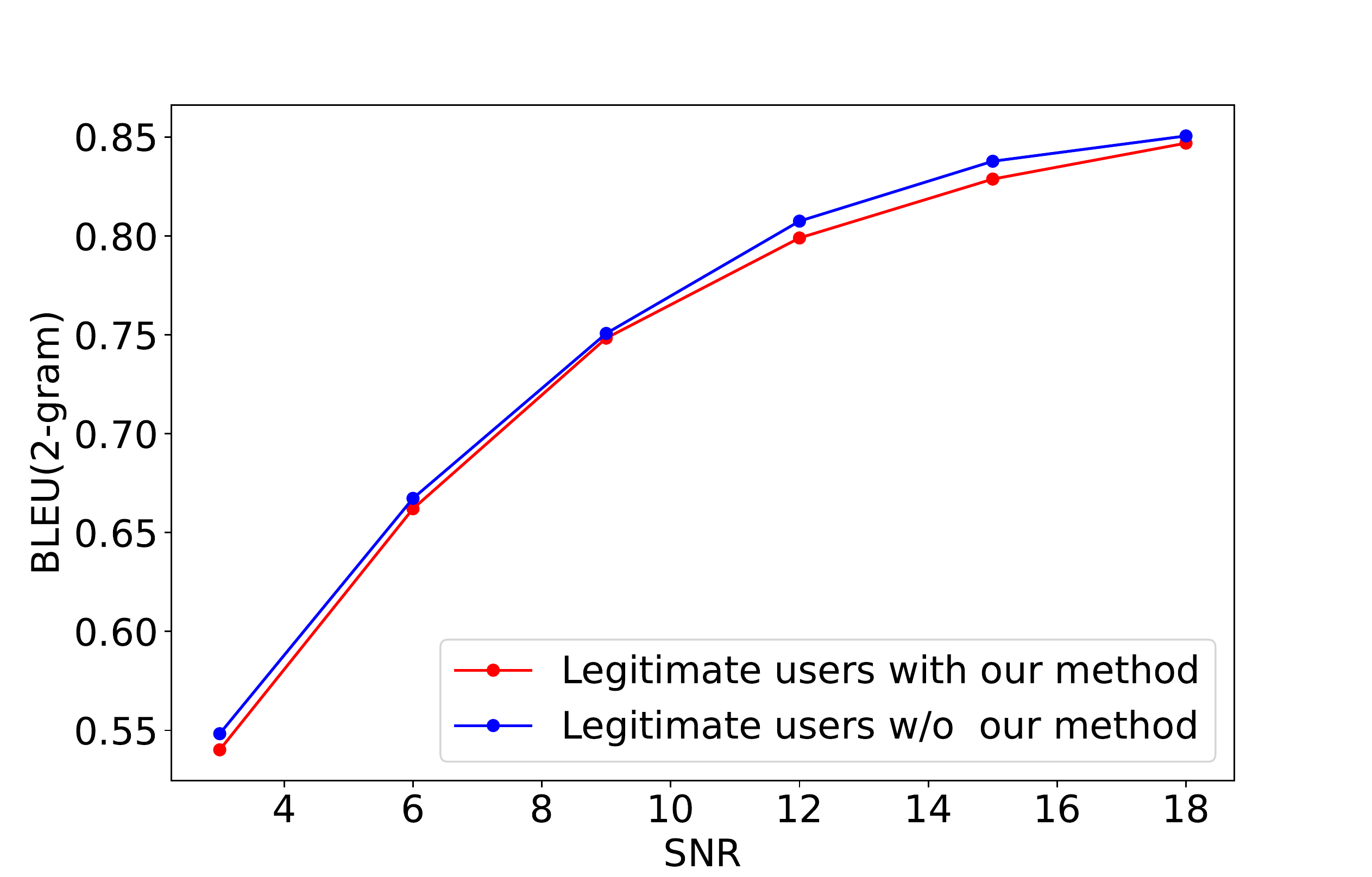}
}
\subfigure[BLEU comparisons under the 3-gram setting.]{
\label{fig:3g}
\includegraphics[width=2.25in,height=1.8in]{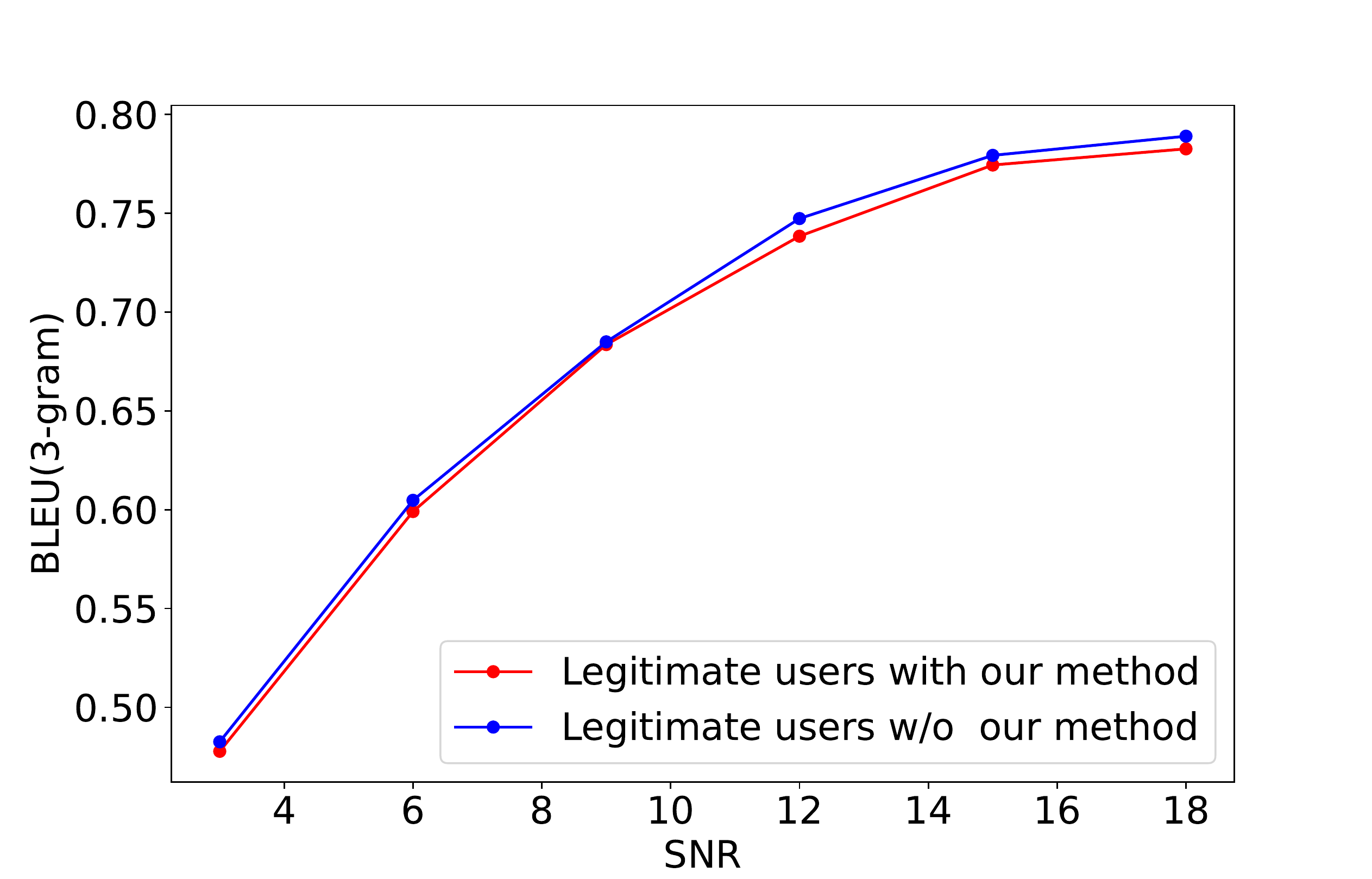}
}
\subfigure[BLEU comparisons under the 4-gram setting.]{
\label{fig:4g}
\includegraphics[width=2.25in,height=1.8in]{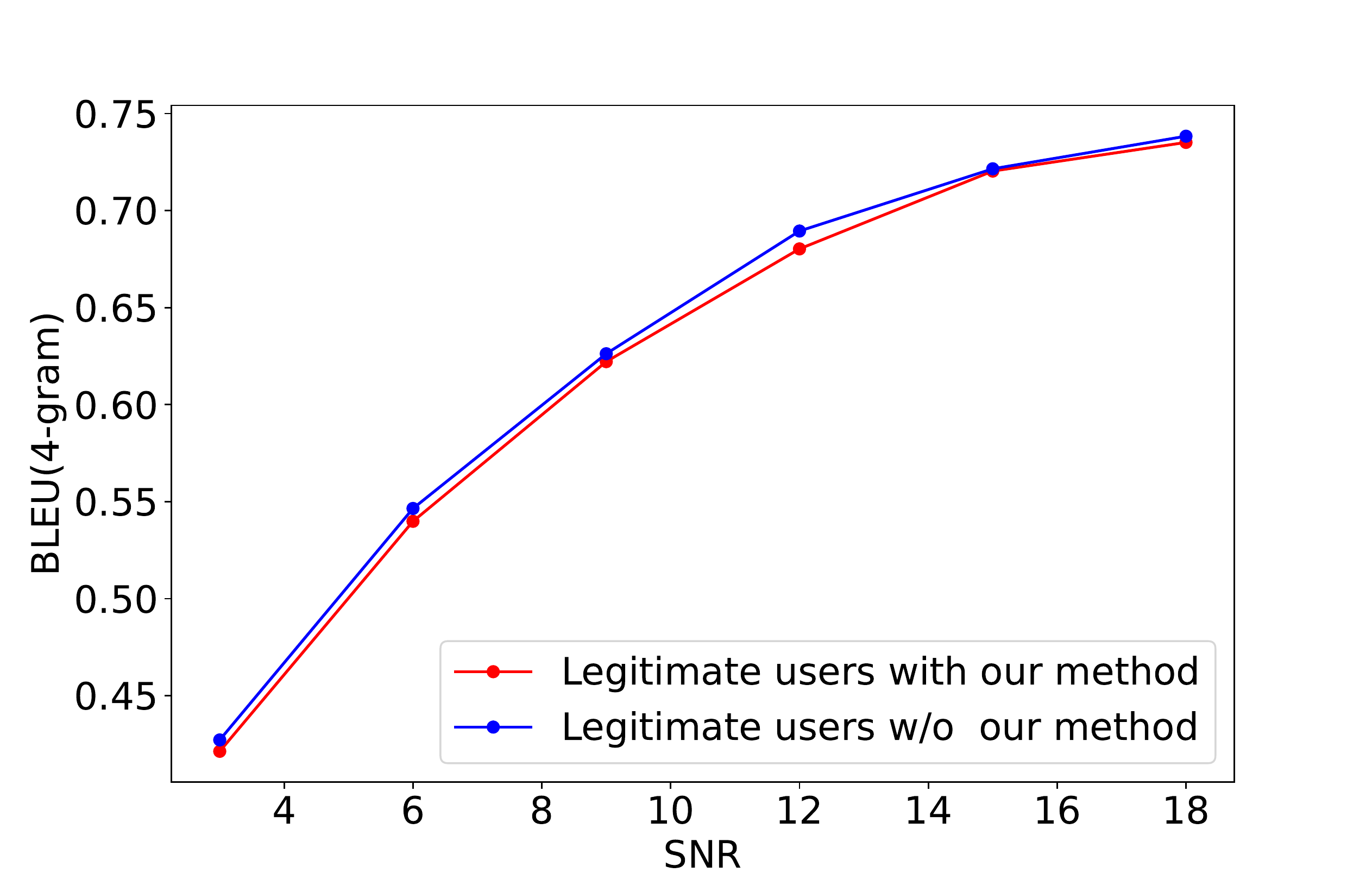}
}
\subfigure[Constellation, SNR = 24 dB]{
\label{fig:Constellation}
\includegraphics[width=2.25in,height=1.8in]{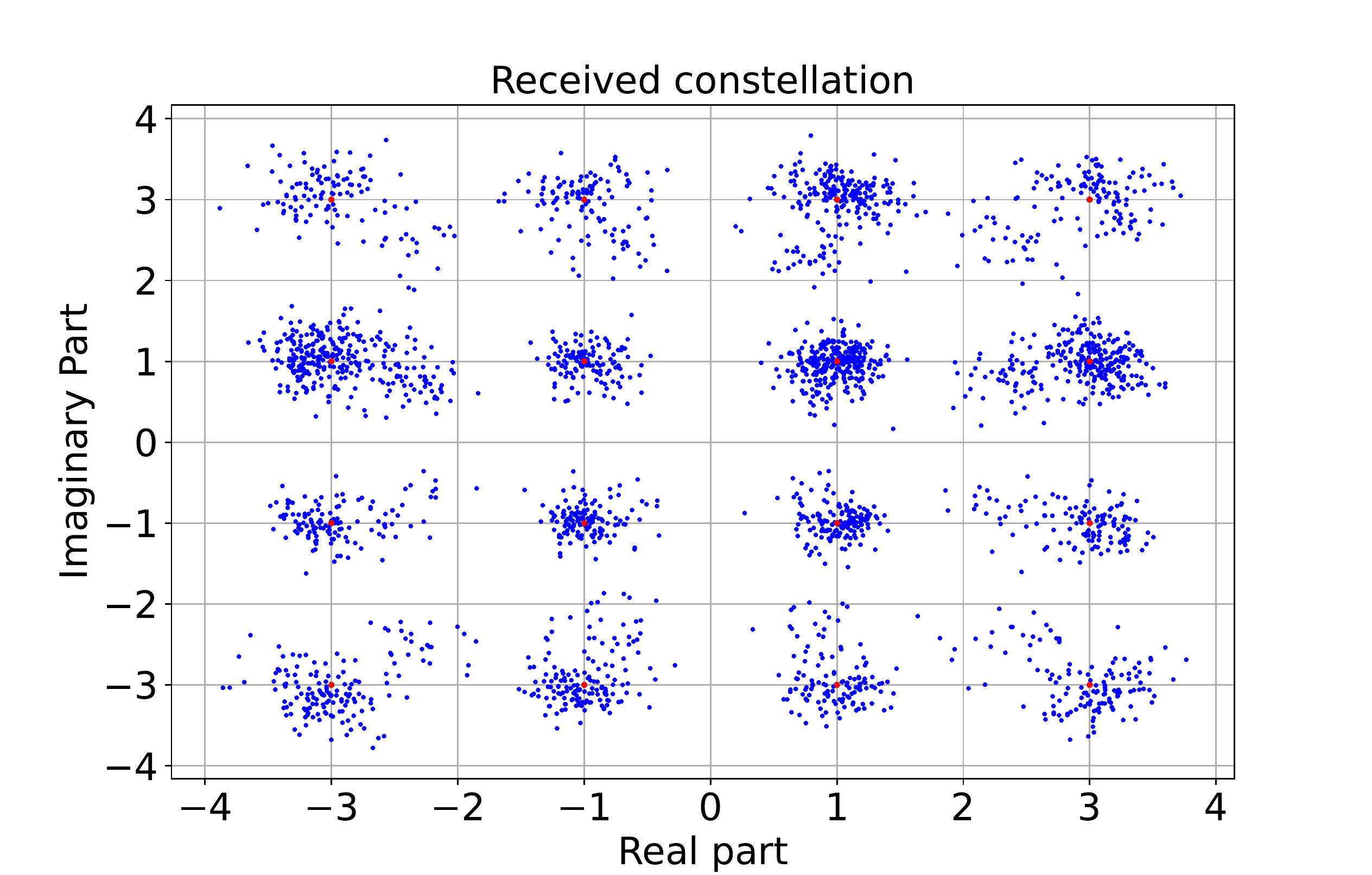}
}
\caption{Effect of our method. (a) the encryption/obfuscation method is transparent to legitimate users in terms of BER, while the BER for an eavesdropper will be significantly degraded. (b), (c), (d) and (e) show the comparisons of BLEU scores under various $N$-gram settings, and the scores with and without our method are almost the same to each other. (f) shows that the semantic data can be correctly demodulated with our encryption/obfuscation method for legitimate users.}
\end{center}
\vspace{-4mm}
\end{figure*}

We employ bit error rate (BER) as our performance metric, which indicates the number of bit errors divided by the total number of transferred bits. Fig. \ref{fig:BER} shows the comparisons of BER curves under various signal-noise-ratio (SNR) for different signals, including the original signal, the one decrypted by legitimate users with our methods and the one decrypted by eavesdroppers. We observe that BER scores under first two settings are almost the same to each other. In contrast, the BER of eavesdroppers with our decryption/deobfuscation will sharply increase to 0.5. The results indicate that eavesdroppers are unable to obtain the semantic conveyed in the DLSC system, and our method will not distort semantic transmission for legitimate users. 

\subsection{Comparisons of BLEU scores}
Fig. \ref{fig:1g}-(e) show the comparisons of BLEU scores with/without our encryption method under $1$-gram, $2$-gram, $3$-gram and $4$-gram settings. For each gram setting, we observe that BLEU scores with and without encryption are slightly different, showing that our methods will not affect semantic interpretations at the receiver. The two modules can be considered as transparency for the end-to-end data transmission.  
\subsection{Effect on OFDM Modulation}
We also demonstrate the effectiveness of our method on OFDM modulation and demodulation during reception with the OFDM constellation map under 16QAM\cite{b3}. Fig. \ref{fig:Constellation} shows that the semantic data can be correctly demodulated with our encryption/obfuscation method.

\section{Security Experiments}
In this section, we quantify the computational overhead of an eavesdropper using the search space of brute-force cracking attacks for the DLSC system. Such a measurement is widely used for security analysis of upper layer encryption schemes.


\subsection{Attack on the Ciphertext}
For such a setting, an eavesdropper only has the knowledge of the encryption algorithm and ciphertext. To attack the DLSC system equipped with our encryption/obfuscation modules, an eavesdropper needs an exhaustive search to generate keys by enumerating all SKey/subcarrier combination until an intelligible translation of the ciphertext to the plaintext is obtained. 

We consider a scenario where an eavesdropper attacks against our proposed encryption scheme. In such a scenario, the eavesdropper first needs to decide the location of the dummy data, and then carries out an exhaustive key search for the decryption. We use the previous method \cite{b1} as our baseline configured with $s$ OFDM symbols and $k$ subcarriers dummy data. We express the parameters as a pair [$s$, $k$]. Assuming that the eavesdropper already knows the pair [$s$, $k$], the search space $ss_{dummy\_location}$ for the dummy data subcarrier position of a data unit can be formulated as:

\vspace{-3mm}
\begin{equation}
ss_{dummy\_location} = C_{s \times N_{d}}^{k}
\label{eq:ssdlocation1}
\end{equation}
where $N_d$ refers to the number of OFDM subcarriers and here $C$ indicates the combination operation. $N_d$ is configured as 64. The search space of dummy location for our method can be expressed as:  
\vspace{-4mm}
\begin{equation}
ss'_{dummy\_location} = \sum_{s=1}^{s_{max}}\sum_{k=1}^{k_{max}} C_{s \times N_{d}}^{k} \,\, (s \times N_d > k)
\label{eq:ssdlocation2}
\end{equation}
where $s_{max}$ and $k_{max}$ are upper bounds of $s$ and $k$, respectively. We set $s_{max}$ and $k_{max}$ as 4 and 10 respectively in experiments. Since the subcarrier positions of the dummy data vary from each data unit, the total search space $ss_{data}$ of dummy data subcarrier in our approach can be expressed as 

\begin{equation}
\begin{aligned}
ss_{data}=(ss_{dummy\_location})^{N_{unit}} = \\(\sum_{s=1}^{s_{max}}\sum_{k=1}^{k_{max}} C_{N_{d}s}^{k})^{N_{unit}}\,\,\,\,(s \times N_d > k)
\end{aligned}
\label{eq:ssdata}
\end{equation} 
where $N_{unit}$ denotes the number of data unit. 

From the above Equation \ref{eq:ssdlocation1} to \ref{eq:ssdata}, we can draw a conclusion that the search space of dummy data in our method is much larger than the one of the baseline, as the values of $s$ and $k$ are dynamic, which are bounded by $s_{max}$ and $k_{max}$. Therefore, our encryption scheme is more secure for semantic protection in DLSC systems.

\subsection{Attacks on the Key Generator}

\begin{figure}[htbp]
\centerline{\includegraphics[width=1\linewidth]{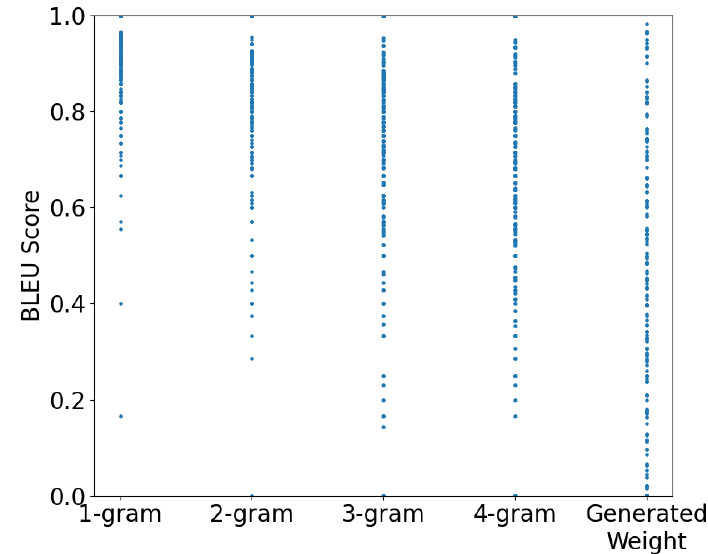}}
\caption{BLEU scores distribution under various $N$-gram settings. The scores under a specific $N$-gram are concentrated and hence the stochastic characteristic for producing keys may be insufficient, while the distribution of weighted sum is more random.}
\vspace{-6mm}
\label{fig:BS}
\end{figure}

\noindent
\textbf{The Randomness of our SKey:}
We generate the semantic key SKey by exploring BLEU scores. Fig. \ref{fig:BS} shows that the BLEU scores under a specific $N$-gram are concentrated and hence the stochastic characteristic for producing keys may be insufficient. We therefore alternatively rely on the weighted sum of four BLEU scores to craft the semantic keys, as we observe that the weights of these scores are more random. The search space of these four weights can be expressed as follows. 
\begin{equation}
ss_{weight}=(2^{L_{weight}})^{4}
\end{equation}
where L$_{weight}$ indicates the number of bits per weight. The search space for the SKey $ss_{S\!K\!ey}$ can be derived by:
\begin{equation}
ss_{S\!K\!ey}=2^{L_{S\!K\!ey}}
\end{equation}
where $L_{S\!K\!ey}$ refers to the length of SKey. Therefore, we can draw a conclusion the search space $ss_{S\!K\!ey}$ is much larger than the search space of generating semantic key under a single $N$-gram, inducing more randomness to secure the semantic communications.

\noindent
\textbf{The Search Space of our Key Generator:}
To attack our key generator, an eavesdropper needs to enumerate all possible keys in $L_{seedkey}$ bits. The corresponding search space is given as follows.
\vspace{-1mm}
\begin{equation}
ss_{seedkey}=2^{L_{seedkey}}
\end{equation}

Assume we use the previous method \cite{b1} to protect the semantic data, the search space of its seed key is given as:
\vspace{-3mm}
\begin{equation}
ss'_{seedkey}=s\times k \times 2^{L_{seedkey}}
\end{equation}
Regarding our encryption method, an eavesdropper needs to obtain the seed key and the value of [$s$, $k$] of each data unit. For each seed key, the eavesdropper requires up to $(s_{max}\times k_{max})^{N_{unit}}$ times attempts for an exhaustive search, and the total search space $ss''_{seedkey}$ can be given as:
\vspace{-1.3mm}
\begin{equation}
ss''_{seedkey}=(s_{max}\times k_{max})^{N_{unit}} \times 2^{L_{seedkey}}.
\end{equation}
Compared with the previous method \cite{b1} that fixed $s$ and $k$, the two parameters in our obfuscation approach are computed by the key stream and vary from each group of OFDM symbols. The total search space for cracking our method can be expressed as:
\vspace{-1mm}
\begin{equation}
\begin{aligned}
ss_{total}=ss_{data}\times ss''_{seedkey} =\qquad\qquad\qquad\qquad\qquad\quad \\
(\sum_{s=1}^{s_{max}}\sum_{k=1}^{k_{max}} C_{N_{d}s}^{k})^{N_{unit}}\times
(s_{max}\times k_{max})^{N_{unit}} \times 2^{L_{seedkey}}
\\ \,\,\,\,(s \times N_d > k).
\end{aligned}
\end{equation}
Therefore, the search space of our encryption scheme $ss''_{seedkey}$ is much larger than that of the original solution. 


\subsection{Key Generation in Static Environments}
For the physical layer key generation, wireless channels are naturally assumed to be dynamic with sufficient randomness, and hence the key generation rate will tend to decrease to ultra-low or even zero in static fading environment. By combing our SKey with traditional PLK, we are able to automatically induce the randomness of the environment for semantic encryption and obfuscation with a large search space $ss_{total}$, leading to high security level of the system. 

\section{Conclusion}
This paper presents a novel physical layer semantic encryption scheme by exploring the randomness of bilingual evaluation understudy (BLEU) scores
in the field of machine translation, and additionally introduces
a novel subcarrier-level semantic obfuscation mechanism to future secure the communications. By exploring the semantic derivation, we are able to encrypt the system with physical layer semantic key in static fading wireless environments. We conduct experiments to verify the effectiveness of our encryption and obfuscation method, and we also provide quantitative analysis to show that the search space of our method is much larger compared to the previous work. In the future, we would like to extend our method to a real test bed.

\section{Acknowledgments}
This work was supported in part by the National Key R\&D Program of China (No. 2022YFB2902200), in part by the National Natural Science Foundation of China (61971066), and in part by the research foundation of Ministry of Education and China Mobile under Grant MCM20180101.

\end{document}